\documentclass[12pt]{article}

\usepackage{graphicx}

\title{STOCHASTIC CORRELATION MODEL \\
OF GALACTIC BULGE VELOCITY DISPERSIONS AND CENTRAL BLACK HOLES MASSES}

\author{V. I. Dokuchaev\thanks{dokuchaev@inr.npd.ac.ru},
        ~Yu. N. Eroshenko\thanks{erosh@ns.ufn.ru} \\
{\small\sl
Institute for Nuclear Research of the Russian Academy of Sciences, Moscow}}

\begin{document}
\date{}
\maketitle

\begin{abstract}
We consider the cosmological model in which a part of the Universe  $\Omega_h\sim10^{-5}$ is
in the form of primordial black holes with mass $\sim10^5M_{\odot}$.
These primordial black holes  would be centers for growing protogalaxies which experienced multiple
mergers with ordinary galaxies. This process of galaxies formation is accompanied by the merging of
central black holes in the galactic nuclei. It is shown that recently discovered correlations between
the central black holes and bulges of galaxies are naturally reproduced in this scenario.
\end{abstract}

\section{Introduction}

Recent observations demonstrate that no less than 20\% of regular
galaxies contain the supermassive black holes (SMBHs) in their nuclei
(Kormendy and Richstone, 1995; Ho, 1998). Different schemes were proposed
for the origin of the central SMBHs due to: (i) gravitational instability
and collapse of the supermassive star (e.~g. Gurevich and Zybin, 1990;
Lipunova, 1997); (ii) collapse of the dynamically evolving dense stellar
cluster (e.~g. Rees, 1984; Dokuchaev, 1991) or (iii) collapse of the
central part of the massive gaseous disk (e.~g. Eisenstein and Loeb,
1995).  In all these scenarios the SMBHs are formed deep inside of the
gravitational potential well of the galactic or protogalalctic nuclei.
The quite specific possibility is the formation of primordial SMBHs in
the early Universe (Zel'dovich and Novikov, 1967;  Carr, 1975).

There are observed three types of correlations between the mass of the central SMBH $M_{BH}$ in the galactic nucleus and (i) stellar mass of galactic bulge $M_b$  (Kormendy and Richstone, 1995; Ho, 1998), $M_{BH}\simeq(0{.}003-0{.}006)M_b$;
(ii) bulge luminosity $L_B$ (Richstone, 1998), $M_{BH}/L_B\simeq10^{-2}M_{\odot}/L_{\odot}$, and
(iii) velocity dispersion $\sigma_e$ at the bulge half-optical-radius (Gebhardt et al., 2000):
\begin{equation}
M_{BH}=1.2(\pm0.2)\times10^8\left(
\frac{\sigma_e}{200\mbox{~km/s}}\right)
^{3.75(\pm0.3)}M_{\odot}.\label{korsig}
\end{equation}
Correlations of types (i) and (ii) now are feebly marked and there is large data scattering with respect to the mean values of $M_{BH}$. So recent observations provide only consistency  between (i) and (ii) in spite of the absence of linear dependence between  $L_B$ and $M_b$. On the other hand  the type (iii) correlation is more definite.

The origin of discussed correlations is quite uncertain. The simplest assumption that the growth of the SMBH mass depends on the bulge processes meets with the problem of different scales: the galactic bulge  scale is a few kpc, whereas the linear scale
of accretion disk around of SMBH is much less than $1$~pc. Some deterministic mechanism is needed for huge mass transfer from bulge to its innermost part.
Silk and Rees (1998) proposed a feasible solution of this problem by considering the
self-adjusting accretion flow from bulge under influence of radiation pressure on the early quasar phase of galactic evolution. However this model predicts the definite type of correlation, $M_{BH}\propto\sigma_e^5$, with the power index differing from observed one.

In this paper we explore the alternative approach by supposing that observed correlations are stochastic in origin. Our basic assumption is the existence
in the Universe of pre-galactic population of black holes with masses $M_h\sim10^5M_{\odot}$ near the recombination time. The similar hypothesis
of the existence of primordial massive black hole population was used by Fukugita and Turner (1996) for interpretation of quasar evolution.

The supposed primordial BHs are mixed with dark matter due to there cosmological origin. So the total mass of these BHs in any galaxy $\sum M_h$ would be proportional to galactic dark matter halo mass $M$. As a result the correlation $\sum M_h\propto M$ is primary in this model, but the aforementioned observed correlations $M_{BH}\propto M_b$, $L$ and $\sigma_e^\alpha$ would be secondary and approximate
in origin due to complicated process of galactic formation. We show below that requested relation $M_{BH}\propto\sigma_e^\alpha$ with a power index $\alpha$ near to observed one follows from our primary relation $\sum M_h\propto M$ with the value of  $\sigma_e$ determined by the dark matter halo mass $M$. The considered model reproduces also the correlation $M_{BH}\propto M_b$ but with a less accuracy.

The other necessary requirement of our model is multiple merging of primordial BHs with mass $M_h$ into the one SMBH with mass $M_{BH}$ during the Hubble time. In  Section~\ref{merge} we validate this requirement. It is known that for a single
BH with mass $M_h\ll10^7M_{\odot}$ the dynamical friction in the galactic halo is ineffective (Valtaoja and Valtonen 1989). Nevertheless for the early formed primordial BHs it is possible the  process of dark matter ``secondary accretion''.
As a result the primordial BHs would be ``enveloped'' by the dark matter halo with a mass of a typical dwarf galaxy and a steep density profile, $\rho\propto r^{-9/4}$.
Indeed, the gravitationally bound objects formed at red-shifts $z\sim10$ from the density fluctuations $\delta\sim10^{-3}$ (at the recombination epoch). In the uniform Universe the BH with mass $M_h\sim10^5M_{\odot}$ produce this fluctuation inside the sphere containing the total mass $M_h/\delta\sim10^8M_{\odot}$. We will call this combined spherical volume ``BH~$+$halo'' by ``induced halo''. The characteristics of these induced halo is defined in Sections~\ref{second} and \ref{okonch}. Our assumption of multiple merging of primordial BHs may be violated in the galaxies of late Hubble types. In fact observations (Zalucci et al., 1998) demonstrate that masses of the central BHs in these galaxies are less than in E and S0 ones and correlation (i)---(iii) are absent.

Throughout this paper we consider for simplicity the flat cosmological model without the cosmological constant. We use index ``i'' for the quantities at the moment
$t_i\simeq6\cdot10^{10}$~s of transition from radiation dominated to matter (dust) dominated phase. Respectively index ``0'' is used for quantities at the recent moment of time $t_0$. We call all BHs existing at nearby the recombination epoch by primordial ones. The other possible mechanisms of BH formation do not influence our results except the Section~\ref{pbhsec}. We use term ``bulge'' as for elliptical
galaxies and for central spheroidal parts of spiral galaxies which reminiscent the dwarf ellipticals.

\section{Primordial Black Holes}
\label{pbhsec}

Noncompact objects (NCOs) of mass  $\sim(0.1\div1)M_{\odot}$, consisting of weakly interacting nonbaryonic dark matter particles $\sim(0.1\div1)M_{\odot}$ like neutralino were proposed by Gurevich et al. (1997) for the explanation of microlensing events in Large Magellanic Clouds. The hypothesized NCOs or neutralino stars are originated from the cosmological fluctuations with a narrow sharp maximum $\sim1$ in the spectrum at some small scale. In addition to neutralino stars the same maximum in the spectrum of cosmological fluctuations produce also the massive primordial BHs with mass $\sim10^5M_{\odot}$ (Dokuchaev and Eroshenko, 2001). So the hypothesized dark matter NCO and primordial BHs may be indirectly connected through their common origin from the same cosmological fluctuations. The spectrum with a sharp maximum at some scale arises in some inflation models (Starobinsky, 1992; Ivanov et al., 1994; Yokoyama, 1995; Garcia-Bellido, 1996). At the same time the spectrum beyond the maximum may be of the standard Harrison-Zel'dovich form and reproduce the usual scenario of large-scale structure formation in the galactic distribution.

Adiabatic density fluctuations of the matter on the scale less than the horizon grow logarithmicallly  in time during the radiation domination epoch. These fluctuations with amplitude $\delta_i\sim1$ at the moment $t_i$ corresponds to radiation density fluctuations $\sim0.05$  at the moment $t_h$ when disturbed region enter into the horizon (Dokuchaev and Eroshenko, 2001). So large fluctuations at radiation dominated epoch produce BHs near the time $t_h$ (Zel'dovich and Novikov, 1967; Carr, 1975).

Formation of BHs takes place on the tail of the Gaussian fluctuation distribution,
whereas the main part of NCO produced from the r.m.s. fluctuations.
Therefore only small part of fluctuations which result in formation of NCO may produce BHs at the radiation domination epoch. In other words due to the large threshold of BH formation the major part of fluctuations do not collapse into the BHs and evolve continuously up to the end of radiation domination epoch.
At the radiation domination epoch the mass in the comoving volume varies as $M(t) = M_x a(t_i)/a(t)$, where the scale-factor of the Universe $a\propto t^{1/2}$
and $M_x$ is the comoving mass at the moment of transition to the matter domination.
The mass $M_x$ equals approximately to the mass of nonrelativistic matter inside the fluctuation, i.~e. the mass of NCO which may be formed from this fluctuation.
On the other hand
\begin{equation}
M(t)=\frac{4\pi}{3}(2ct)^3\rho(t),
\end{equation}
where $\rho(t)=3/32\pi Gt^2$. From these relations for $M(t)$ we find the time $t_h$ and mass $M_h$ of BH formation in dependence of the mass of NCO:
\begin{equation}
t_h=(M_x G)^{2/3}c^{-2}t_i^{1/3}=1.1\left(\frac{M_x}{M_{\odot}}\right)^{2/3}
\left(\frac{t_i}{6\cdot10^{10}\mbox{~c}}\right)^{1/3}\mbox{~s},
\label{tm}
\end{equation}
\begin{equation}
M_h=cM_x^{2/3}G^{-1/3}t_i^{1/3}=2.3\cdot10^5\left(\frac{M_x}{M_{\odot}}\right)^{2/3}
\left(\frac{t_i}{6\cdot10^{10}\mbox{~c}}\right)^{1/3}M_{\odot}.
\label{mhmx}
\end{equation}
In our model the NCOs add BHs are originated from the fluctuations of the same type but at different moment of time. Whereas the  strong difference in masses of NCO add BH is accounted by the mass variation in time inside the fixed comoving volume:
the mass of radiation at the radiation domination epoch (e.~g. at moment $t\sim1$~s)
far exceeds the cold dark matter mass at the matter domination epoch in the same comoving volume.

There are definite astrophysical limitations on the number and mass of primordial BHs:
\begin{enumerate}
  \item From the Universe age limit it follows that fraction of BHs $\Omega_h\le1$.
Additionally the PBHs with $\Omega_h\sim1$ would disturb the background microwave
spectrum if they formed after the time $\sim1$~s of electron-positron pair annihilations (Carr, 1975).
  \item The possibility of tidal destruction of globular clusters by the primordial BHs gives the BH mass limit $M_h\le10^4M_{\odot}$ if these BHs provide the major contribution to the dark matter (Moore, 1993).
  \item The model-dependent calculations of the contribution from matter accretion by PBHs at pre-galactic and recent epoch into background radiation give approximately $\Omega_h\le10^{-3}\div10^{-1}$ for $M_h\sim10^5M_{\odot}$ (Carr, 1979).
  \item From the absence of reliable events of gamma-ray bursts (GRBs) lensing it is obtained the limit $\Omega_h<0.1$ for PBHs in the intergalactic medium with masses $10^5M_{\odot}<M_h<10^9M_{\odot}$ (Nemiroff et al., 2001). An even more stringent limit $\Omega_h<0.01$ for $10^6M_{\odot}<M_h<10^8M_{\odot}$ is obtained from VLBI observations of the lensing of compact radio-sources (Wilkinson et al., 2001).
\end{enumerate}
In this paper we consider the case $\Omega_h\sim10^{-5}$ in accordance with all aforementioned limits.

\section{Formation of Induced Halo}
\label{second}
A massive induced halo (IH) or a heavy dark matter envelope around the PBH is formed due to dark matter accretion in the expanding Universe. Spherical matter accretion onto the compact object (SMBH or galaxy) produces the stationary matter density profile of the form $\rho\propto r^{-9/4}$ (e.~g. Ryan, 1972; Gunn, 1977). Here we reproduce the similar calculations in the form suitable for binding the mass of PBH with a mass of gravitationally trapped matter and a corresponding red-shift.

The matter in the spherical layer of radius $r$ around the PBH in the uniform Universe exerts the acceleration
\begin{equation}
\frac{d^2r}{dt^2}=-\frac{G}{r^2}
(M_h+\frac{4\pi}{3}r_i^3\rho_i),
\label{d2rdt}
\end{equation}
where $r_i$ and $\rho_i=4.4\cdot10^{-18}(h/0.6)^8\mbox{~g~cm}^{-3}$
is correspondingly the radius of the spherical layer and the Universe density
at the moment $t_i$ when the strong growth of fluctuations is beginning.
and $h$ is the Hubble constant in units of $100$~km~s$^{-1}$~Mpc$^{-1}$.
Let us multiply (\ref{d2rdt}) on $db/dt$ and integrate it by using normalization $r(t)=r_ib(r_i,t)$ and initial condition $b(t_i)=1$, $\dot b(t_i)=H(t_i)b$.
After using an additional normalization $\tau=(t-t_i)\alpha^{1/2}$, where
\begin{equation}
\alpha=\left(\frac{2GM_h}{r_i^3}+\frac{8\pi G}{3}\rho_i\right),
\label{norm}
\end{equation}
the integrated (\ref{d2rdt}) takes the form:
\begin{equation}
b(db/d\tau)^2=1-bE,
\label{bdb}
\end{equation}
with  $E\equiv(\alpha-H^2(t_i))/\alpha$. Equation of this type admit an exact solution (Saslaw, 1989). From this solution we find the expansion termination moment of
the considered layer:
\begin{equation}
t_{s}\approx\frac{3\pi}{4}\left(\frac{M_s}{M_h}\right)^{3/2}t_i.
\label{ts}
\end{equation}
Here $M_s=4\pi\rho_ir_i^3/3$ is the mass inside the layer with the exception of
central PBH mass. At the moment of expansion termination $b(t_s)=1/E$ and subsequently the layer is contracted up to the moment $t_{col}\approx2t_s$. From (\ref{ts}) by using relation $t=t_0/(1+z)^{3/2}$ for the flat Universe without the cosmological constant  we obtain
\begin{equation}
z_{col}\approx0.36z_i\frac{M_h}{M_s}-1\approx
2\cdot10^3\frac{M_h}{M_s}-1,
\label{zs}
\end{equation}
where $z_i=2.4\cdot10^4h^2$.

We will suppose that spherical layer is detached from cosmological expansion and virialized and after its expansion termination and a subsequent contraction from radius $r_ib(r_i,t_s)$ to radius $r_{col}=r_ib(r_i,t_s)/2$. For this virialization radius of the IH around the central PBH we find
\begin{equation}
r_{col}=\frac{r_i}{2E}=\frac{M_s^{4/3}}{2M_h}\left(\frac{3}{4\pi\rho_i}\right)^{1/3}
=0.45\left(\frac{M_h}{10^5M_{\odot}}\right)^{1/3}
\left(\frac{15}{1+z_{col}}\right)^{4/3}\mbox{~kpc}.
\label{rcol}
\end{equation}
Now we obtain the requested relation for the mass and radius of IH
\begin{eqnarray}
M_s(r_{col})&=&(2M_hr_{col})^{3/4}\left(\frac{4\pi\rho_i}{3}\right)^{1/4} \nonumber\\
&=&3.7\cdot10^7\left(\frac{M_h}{10^5M_{\odot}}\right)^{3/4}
\left(\frac{r_{col}}{1\mbox{~kpc}}\right)^{3/4}\left(\frac{h}{0.6}\right)^{2}
M_{\odot}
\label{mgrcol}
\end{eqnarray}
and density distribution in the IH
\begin{eqnarray}
\rho(r)&=&\frac{1}{4\pi r^2}\frac{dM_s(r)}{dr} \nonumber\\
&=&1.7\cdot10^{-25}\left(\frac{M_h}{10^5M_{\odot}}\right)^{3/4}
\left(\frac{r}{1\mbox{~kpc}}\right)^{-9/4}\left(\frac{h}{0.6}\right)^{2}
\mbox{ g~cm}^{-3}.
\label{rho}
\end{eqnarray}

\section{Protogalaxies Growth Termination}
\label{okonch}

The growth of IH terminates at the epoch of nonlinear growth of ambient density fluctuations with mass $M$ and radius $R$ producing the same gravitational acceleration:
\begin{equation}
\frac{GM_s(r_{col})}{(2r_{col})^2}=\frac{GM}{(2R)^2}, \label{ggg}
\end{equation}
where radiuses $2r_{col}$ and $2R$ correspond to the moment of expansion termination.
At this condition the capture of new spherical layers is shutting down (see. (\ref{rcol})).

The r.m.s. fluctuation in the mass scale $M$ is defined by expression (Bardeen et al., 1986)
\begin{equation}
\sigma^2(M,z)= \frac{1}{2\pi^2(1+z)^2}
\int dkk^2P_0(k)W^2(k,M).
\end{equation}
Here $W(k,M)$ is a filtering function and $P(k)$ is a power spectrum:
\begin{equation}
\langle\delta^*_{k}\delta_{k'}
\rangle=(2\pi)^3P(k)\delta_D^{(3)}(k-k'), \quad
\delta_{k}=\int\delta(r)e^{ikr}d^3r,
\label{pow}
\end{equation}
where $\delta_D^{(3)}(k-k')$ is the Dirac delta-function and angular brackets denote the ensemble averaging. Let us choose $P_0(k)$ for the model with a cold dark matter
(Bardeen et al., 1986) with the Harrison-Zel'dovich initial spectrum:
\begin{equation}
P_0(k)=\frac{ak\ln^2(1+2.34q)}{(2.34q)^2}
(1+3.89q+(16.1q)^2+(5.46q)^3+(6.71q)^4)^{-1/2},
\label{cdm}
\end{equation}
where $q=k\Omega_d^{-1}h^{-2}$ is written in the comoving coordinates with units Mpc$^{-1}$. For simplicity we put the dark matter density parameter $\Omega_d=1$.
Normalization constant $a$ is chosen from the condition $\sigma_0=1$ at the scale
$8h^{-1}$~Mpc.
The condition for object formation with the mass $M$ at the red-shift $z_{col}$ is
\begin{equation}
\sigma(M,z_{col})=\delta_c\label{sig1},
\end{equation}
where for the model of spherical contraction  $\delta_c=3(3\pi/2)^{2/3}/5$ (White, 1994). Radius of ordinary protogalaxy (without the PBH) after its cosmological expansion termination and virialization is (Saslaw, 1989; White, 1994)
\begin{equation}
R=\frac{3R_i}{10\delta_i}, \quad R_i=\left(\frac{3M}{4\pi\rho_i} \right)^{1/3}.
\label{rxxxx}
\end{equation}
By taking into account (\ref{rcol}), (\ref{sig1}), (\ref{rxxxx}), and putting
$\delta_i=\sigma_i(M)$ the condition (\ref{ggg}) can be rewritten in th form:
\begin{equation}
M=\left(\frac{3\pi}{2}\right)^{10/3}\left(\frac{3}{5}\right)^{6}
\frac{(1+z_i)}{(1+z_{col})}M_h\delta_c^{-6}.
\label{k6}
\end{equation}
It follows that $M_s=M$, because in (\ref{rcol}) and (\ref{rxxxx})
it is supposed that IHs and ordinary protogalaxies have the same coefficient of nonlinear contraction equals to $0.5$. We solve numerically the system of equations
(\ref{sig1}) and (\ref{k6}) relative to independent variables $z_{col}$ and $M$ with the $M_h$ as parameter.
See Fig.~1 for calculated relations for $z_{col}(M_h)$ and $M(M_h)=M_s(M_h)$.
For  $M_h=2.3\cdot10^5M_{\odot}$ we find $z_{col}=8.8$ and
$M=7.2\cdot10^7M_{\odot}$. As a result up to epoch $z=8.8$ the PBHs with mass $M_h=2.3\cdot10^5M_{\odot}$ had time to capture an additional mass which $\sim300$  times exceeds the PBH mass.

\section{Merging of Black Holes}
\label{merge}
In the preceding Chapter we demonstrate that massive IHs with mass $M_s=7.2\cdot10^7M_{\odot}$ are formed around the PBHs. These IHs are massive enough to sink down to the galactic center during the Hubble time under influence of dynamical friction. Though the fate of nested PBHs inside the central parsec of the host galaxy is rather uncertain.
Valtaoja and Valtonen (1989) considered interaction of central BHs after merging of galaxies. The late phase of two BHs merging in the galactic center depends on many factors (Menou et al., 2001). Without the detailed elaboration we will follow to
Menou et al. ( 2001) by supposing that multiple PBHs merge into a single SMBH during the Hubble time.

Notice that density (\ref{rho}) strongly grows towards the center and
smoothed out only at the distance $r_h$ where $M_s=M_h$ according (\ref{mgrcol}).
For $M_h=2.3\cdot10^5M_{\odot}$ we obtain $r_h\sim1$~pc and density
$\rho\sim10^4M_{\odot}$~οκ$^{-3}$. By using Chandrasekhar time for dynamical friction (see e.~g. Saslaw, 1989) of the PBH with mass $M_h=2.3\cdot10^5M_{\odot}$ we obtain the estimation of characteristic time for PBH merging
\begin{equation}
t_f\sim \frac{v^3}{4\pi G^2\Lambda B\rho M_h}\sim 5\cdot10^5\mbox{~yrs},
\end{equation}
where $v\sim (GM_h/r_h)^{1/2}$ is the PBH velocity,
$\Lambda\approx10$, $B\approx0.426$. The PBH of larger mass would be merged even faster. As a result the late phase of PBHs merging lasts very fast, and the probability  simultaneous presence in the galactic nucleus of three or more BHs is low.

Coalescence of PBHs in the galaxies must be accompanied by the strong burst of gravitational radiation. The projected interferometric detector LISA is capable to detect the coalescence events up to the red-shifts$z\sim10$, if BH masses no less than $10^3M_{\odot}$ (Menou et al., 2001). For the calculation of gravitational bursts distribution there is needed the numerical simulation of hierarchical BH and galactic merging. Menou et al., (2001) carried out the corresponding simulation for ordinary galaxies.  A simple estimation of the burst rate from the observable Universe gives\begin{equation}
\dot N_{grav}\sim\frac{4\pi}{3}\frac{N}{t_0}(ct_0)^3~n_g
\sim10 \left(\frac{n_g}{10^{-2}\mbox{Mpc$^{-3}$}}\right)
\left(\frac{t_0}{10^{10}\mbox{yrs}}\right)^2
\left(\frac{N}{100}\right)\mbox{~yrs$^{-1}$},
\end{equation}
where $n_g$ is a mean number density of structured galaxies and
$N$ is a mean number of merging per galaxy. So there is
a principal possibility for the verification of considered model by the LISA detector.

\section{Origin of Correlations}
\label{koritog}

Fluctuation spectrum (\ref{cdm}) in the confined mass region can be approximated by the power law with the effective index $n=d\ln P_0(k)/d\ln k$. According to the formation condition  (\ref{sig1}) for power law spectrum the effective galactic mass formed at red-shift $z$ is (White, 1994)
\begin{equation}
M=M_0(1+z)^{-\frac{6}{n+3}},
\end{equation}
where $M_0=const$. For galactic mass $M=10^{10}M_{\odot}$ we obtain
$n=-2.28$ θ $M_0=2.5\cdot10^{16}M_{\odot}$ and for $M=10^{12}M_{\odot}$ we find
correspondingly $n=-1.98$ θ $M_0=7\cdot10^{14}M_{\odot}$.
Velocity dispersion is estimated as
\begin{equation}
\sigma_e^2\simeq\frac{GM}{R}, \quad
R=\left(\frac{3M}{4\pi\rho(z)}\right)^{1/3}, \quad
\rho(z)=\rho_0(1+z)^3.
\label{line}
\end{equation}
For PBH cosmological density parameter $\Omega_h$ and effective PBH merging in the galaxies the preceding relation gives the final mass of the central BH
\begin{equation}
M_{BH}=\psi\Omega_hM=\psi\Omega_{h}\sigma_e^{\frac{12}{1-n}}
M_0^{-\frac{n+3}{1-n}}\left(\frac{4\pi G^3\rho_0}{3}\right)^{-\frac{2}{1-n}},
\label{bhn}
\end{equation}
where factor $\psi$ is responsible for the possible additional growth of the central BH due to accretion. From (\ref{bhn}) with $h=0.6$ it folows
\begin{equation}
M_{BH}=(1.1\div1.7)\cdot10^8\left(\frac{\psi\Omega_{h}}{10^{-5}}\right)
\left(\frac{\sigma_e}{200\mbox{~km~s}^{-1}}\right)^{(3.66\div4.03)}M_{\odot},
\end{equation}
where pair of coefficients 1.1 and 3.66 corresponds to
$M=10^{10}M_{\odot}$, and respectively pair 1.7 and 4.03 correspond to $M=10^{12}M_{\odot}$. In summary the considered model is in a good agreement
with observation data (\ref{korsig}). The fluctuation spectrum at the galactic scale, $n\approx-2$,
 completely defines the power index $\alpha\approx4$ in the relation $M_{BH}\propto\sigma_e^\alpha$. It is easily verified that for $\Omega_h\sim10^{-5}$ the contribution IHs (there mass is determined in Section~\ref{okonch}) to the total galactic mass is negligible.

Correlation of the form $M_{BH}\propto M_b$ is explained by the common
relation of dark matter mass in galaxies of all types with their total
baryonic (star) mass with the main input from old stars. In the framework
of hierarchical model of galaxy formation the recent galaxies formed from
multiple merging of low mass protogalaxies with early star formation
proceeding. in this case the old star population of spherical subsystem
is  protogalactic in origin.

The part of galactic mass contained in stars becomes a fixed and independent part of the total galactic mass $M$ due to statistical averaging after the large number of protogalactic merging. So the total star mass in the galaxy is $M_b=f_bf_sM$, where
$f_b\approx0.05$ is the baryonic mass fraction of the Universe, $f_s$ is the part of baryons passing into the stars. After sinking down to the galactic center and merging of all PBH the resulting mass of the BH is
\begin{equation}
M_{BH}\simeq\frac{\psi\Omega_h}{f_bf_s}M_b\simeq10^{-3}
\left(\frac{\psi\Omega_h}{10^{-5}}\right)
\left(\frac{f_bf_s}{0.01}\right)^{-1}M_b,
\end{equation}
which with the order of magnitude corresponds to observations
$M_{BH}\simeq(0.003-0.006)M_b$.

The central BH masses in Sa, Sb, Sc galaxies are less in a mean than in E and S0 galaxies (Salucci et al., 1998). In our model this is related with a relatively late formation of Sa, Sb, Sc galaxies when the main part of PBHs do not have enough time to sinking to the galactic center. In particular $\sim10^2$ PBHs of mass $M_h\sim10^5M_{\odot}$ can inhabit our Galaxy.

\section{Conclusion}

We explain the observed correlations between bulge parameters and central BH masses in galaxies by the multiple merging of PBHs of mass $\sim10^5M_{\odot}$ generated in the early Universe. This model predicts the existence of BHs with mass $\sim10^5M_{\odot}$ in beyond the dynamical centers of spiral galaxies and in the intergalactic medium. Possibly one of these BHs was detected by the Chandra Observatory in the galaxy M82 (Kaaret et al., 2000). The observational signature for verification of this model is the gravitational burst after BH merging in the galactic nuclei which may be detected by the projected laser interferometric detectors of the LISA type.

The work was supported in part by the INTAS grant 99--1065 and by Russian Foundation for Basic Research grants 01-02-17829, 00-15-96697 and 00-15-96632. We also acknowledge the referees of this paper for helpful remarks.

\newpage

FIGURE CAPTIONS

\bigskip

Fig. 1. Functions $z_{col}(M_h)$ and $M(M_h)$ from the numerical
calculation of the system of equations (\ref{sig1}) and (\ref{k6}).

\newpage

\begin{figure}
\includegraphics{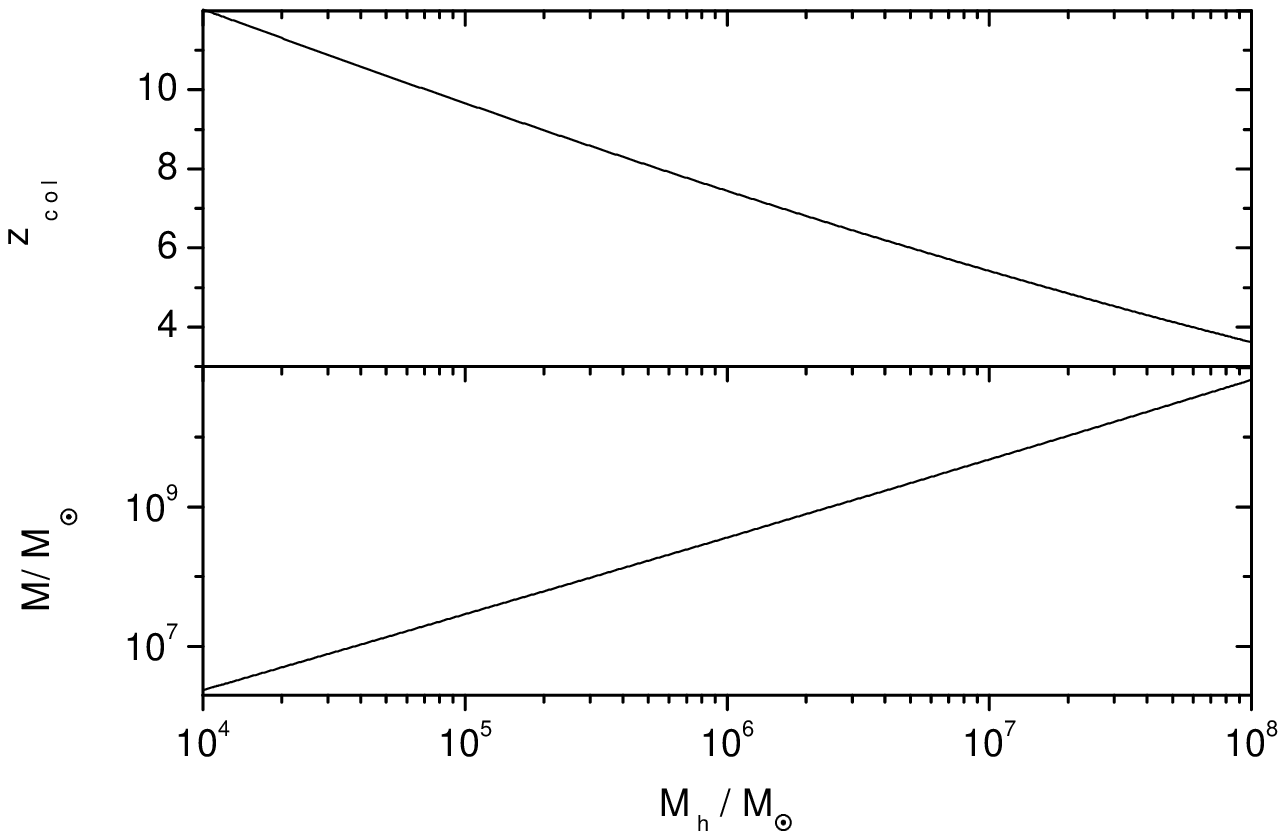}
\label{figgg1}
\end{figure}

\end{document}